\documentclass[runningheads,a4paper]{llncs}

\usepackage{amssymb}
\usepackage{amsmath}
\usepackage{graphicx}
\usepackage{caption}
\usepackage{epstopdf}
\usepackage{url}
\usepackage{csquotes}
\urldef{\mailsa}\path|{marouen.benguebila, johan.thunberg}@uni.lu|    
\newcommand{\keywords}[1]{\par\addvspace\baselineskip
\noindent\keywordname\enspace\ignorespaces#1}

\begin{document}

\mainmatter 

\title{Toward a closed-loop subcutaneous\\
delivery of L-DOPA }

\titlerunning{Closed-loop L-DOPA subcutaneous pump }

\author{Marouen Ben Guebila
\and Johan Thunberg}

\authorrunning{Closed-loop L-DOPA subcutaneous pump}

\institute{Luxembourg Centre for Systems Biomedicine, Campus Belval\\
7, avenue des Hauts-Fourneaux. \\
L-4362 Esch-sur-Alzette, Luxembourg\\
\mailsa\\
}

\toctitle{Closed-loop L-DOPA subcutaneous pump}
\maketitle

\begin{abstract}
L-DOPA has been the gold standard treatment for Parkinson's disease since 50 years. Being the direct biochemical precursor of dopamine, L-DOPA is effectively converted in the brain, but two major phenomena reduce its therapeutic action: i) competition with amino acids in the gut wall and in the blood brain barrier and ii) its fast kinetics (absorption, distribution, metabolism, and elimination). Continuous administration of L-DOPA, such as jejunal pumps, have addressed the issue of fast absorption. Considering a subcutaneous delivery of L-DOPA allows to bypass the gastrointestinal tract and avoid competition with dietary amino acids. Remains the competition at the blood barrier between amino acids and L-DOPA, which we address by proposing a closed-loop controlled, continuous subcutaneous delivery pump. In the proof-of-concept format, the delivery strategy evaluated on comprehensive model of L-DOPA kinetics, holds the promise of improving the treatment of late-stage Parkinson's disease patients.
\end{abstract}

\section{Introduction}
L-DOPA (levodopa) is a naturally occurring precursor of dopamine in the human body \cite{contin2010pharmacokinetics}. The loss of dopaminergic neurons and the subsequent impaired production of dopamine in the nigrostriatal region of the brain is the major cause of pathogenesis and progression of Parkinson's disease \cite{fahn2015levodopa}. The supplementation of L-DOPA is the gold standard in the treatment of Parkinson's disease since 50 years. The efficacy of L-DOPA is hampered by its inherent fast kinetics. After per os administration of a single dose, the peak of the concentration occurs within an hour, after which a decrease in the concentration rapidly follows \cite{contin2010pharmacokinetics}. This kinetic profile leads to an on/off phenomena. The on phenomenon refers to the periods when the symptoms are controlled. During such periods, dyskenesia (uncontrolled movements) might appear while the peak of the concentration exceeds the upper limit of the therapeutic window \cite{nutt1984off}. The off phenomenon, or akinesia, occurs when the concentration of L-DOPA drops below the therapeutic window. Patients report the inability to move and blockage, causing fall injuries \cite{schrag2000dyskinesias}. 
Other factors have been found to contribute to the on/off fluctuations, such as diet, genetic variants, and stress \cite{contin2010pharmacokinetics}.
The continuous intravenous administration of L-DOPA has been
shown to be correlated with lower on/off phenomena \cite{senek2014continuous}, which drove advances in L-DOPA formulation \cite{nutt1984off}. A continuous administration of L-DOPA gel through a jejunal pump, decreased the off-phenomena by 1.9 hours a day \cite{lundqvist2007continuous}. Although, it has been reported that continuous administration exacerbated adverse reactions related to levodopa, mainly the decrease in folate, vitamin B12, and vitamin B6 as well as the increase in homocysteine and methylmalonic acid, which induce neurological and metabolic dysfunctions \cite{rogers2003elevated}.\\ 
Real-time controlled pumps are a new generation of drug delivery devices, which take into account the physiology of patient as well as the drug concentrations to deliver a personalized dose. Such devices allow for fast responses to unwanted changes in order to keep drug levels within a predefined therapeutic window \cite{kovatchev2009silico}. Although subcutaneous blood concentration monitoring of L-DOPA is not yet routinely done, non-invasive techniques have been developed and offer the promise of a fast and reliable measurement of L-DOPA concentrations. Notably, microdialysis, a minimally invasive technique, has been tested and increasingly adopted in clinical practice for Parkinson's disease patients under L-DOPA treatment. Bypassing the small intestine through a subcutaneous infusion improved the bioavailability of L-DOPA \cite{giladi2015pharmacokinetics}. It has been shown that maintaining steady state concentrations of L-DOPA through continuous infusion did not guarantee achieving the clinical objectives, especially with proteic diet \cite{frankel1989effects}.\\
Taken together, (i) the fluctuations in the response to L-DOPA as well as (ii) its gastrointestinal absorption characterized by erratic peaks, (iii) the adverse reactions shown by the uncontrolled continuous jejunal delivery, and (iv) the failure to improve the symptoms through only maintaining a steady concentrations of plasma L-DOPA, make the subcutaneous controlled delivery pump a promising alternative for the treatment of Parkinson's disease.
To this end, a modified molecule, L-DOPA methyl ester, was successfully administered through the subcutaneous route \cite{kleedorfer1991subcutaneous}. The feedback control would ideally come from (i) measuring L-DOPA concentrations in the plasma, (ii) resting tremor (hand shacking) as a surrogate endpoint, and (iii) brain dopamine production.Since the latter measurement needs invasive techniques, it should be reserved for patients subjected to deep brain stimulation using implanted macroelectrodes that allow the measurement of the local field potential (LFP) and the detection of brain power spectrum correlated to on and off phenomena, as shown previously \cite{alonso2006slow}. \\
Here, we computationally model and analyse the behaviour of a novel subcutaneous delivery controlled pump based on comprehensive model of L-DOPA kinetics and dynamics that merges two published models (\cite{chan2005importance} \cite{contin2010pharmacokinetics}), and compare different control strategies (time invariant and continuous). 

\section{Output measurements}\label{sec:output_measurements}
The model describing L-DOPA is a combination of the pharmacokinetics
 \cite{chan2005importance} and the pharmacodynamics \cite{contin2010pharmacokinetics} of the drug and consists of three state variables $X_1(t)$, $X_2(t)$, and $X_3(t)$ that represent L-DOPA concentrations in the plasma, tissues, and the brain respectively. \\
Presently, we will develop on the state variable $X_3(t)$ that is directly correlated to the drug efficacy. It is assumed that the effect of the L-DOPA cannot 
be measured directly, i.e., we cannot measure $X_3(t)$.\\
Instead, $X_3(t)$ can only be observed indirectly from the so called outputs. One choice for an output is the effect corresponding on hand tapping; another choice is the LFP measurements in patients equipped with deep brain stimulation devices; a third alternative is $X_1(t)$, i.e., the plasma concentration. Measuring 
$X_3(t)$ will not be further discussed, but 
the other two -- hand tapping and LFP -- will now 
be addressed. \\

\textbf{Hand tapping}: \\
The concentration $X_3(t)$ can be directly estimated by the number
of hand taps a patient performs in a minute. 
A model of the hand taps was provided in \cite{contin2010pharmacokinetics} and is 
given by $E(t) = (h \circ X_3)(t) = \dfrac{E_{\max}X_3^n(t)}{EC_{50}^n+ X_3^n(t)} + E_0.$
The function $h$ is implicitly parametrized by 
$E_{\max}s^n$, $EC_{50}^n$, $E_0$, and $n$.
$E(t)$ represents the effect. In this situation, it is the number of taps per minute that the patient were asked to do on buzzers separated by a distance of 20 cm. The integer $n$ is the hill coefficient, $E_{\max}$ is the maximum effect, and $EC_{50}$ is the concentration of L-DOPA in the effect compartment that gives $50\%$ of the maximum effect.\\
The number of taps  is measured as the number of times the patient could alternately tap two buttons 20 cm apart in 60 seconds, with the most affected hand \cite{contin2010pharmacokinetics}. The hand shacking is assumed to have a similar profile as the tapping test if measured with an accelerometer or a similar body metric device.\\
It is also assumed that subcutaneous administration of L-DOPA yields the same kinetics as the intravenous route. 
As a motivation for using the effect of hand tapping to 
indirectly measure $X_3(t)$, it should be mentioned that wearable sensors that can measure soft structures and body metrics are an active field of research \cite{o2012artificial}. \\

\textbf{LFP measurements}:\\
In this case, it is assumed that $X_3(t)$ can be approximated via 
LFP in patients equipped with deep brain stimulation devices. Measuring the level of dopamine in the brain is the ultimate efficacy indicator about disease progression and L-DOPA efficacy. It has been reported that the increase of L-DOPA in the brain is correlated with power spectrum modification in the LFP \cite{alonso2006slow}.\\
Dyskinetic patients showed a higher Gamma power in the LFP profile, while akinetic patients have a Beta power profile. The change in the LFP profile has been reported to correlate with a change in dopamine concentration in the brain and dyskinesia \cite{alonso2006slow}. 

\section{A comprehensive model of L-DOPA kinetics: Compact notation}\label{sec:model}
The model introduced can 
be represented by
\begin{align}
\label{eq:sys}
\begin{cases}
\dot{x} & = Ax + Bu, \\
y & = f(x),
\end{cases}
\end{align}
where $x(t) = [X_1(t), X_2(t), X_3(t)]^T \in \mathbb{R}^+$ are the state variables and $u(t) = [u_1(t),u_2(t)]^T $ is defined by
$u_1(t)  = \frac{k_c(t)}{V_1} + \frac{R_{\textnormal{syn}}}{(k_{\textnormal{el}} V_1)} \in \mathbb{R}^+, 
u_2(t)  = \frac{k_e(t)}{V_1} \in \mathbb{R}^+
$ 
with $R_{syn}$ being the endogenous rate of synthesis, $V_1$ the volume of blood compartment, $k_{el}$ the clearance rate, $k_{c}(t)$ and $k_{e}(t)$ are the infusion rates in the central and peripheral compartment respectively and $y(t)$ is the output, i.e., states or functions thereof, which are assumed to be measurable.  The control variable $u(t)$ is introduced to be 
able to write the dynamics of $x$ as a linear time invariant system. In the 
end, it is the rate $k_c(t) = V_1\left(u_1(t) - \frac{R_{\textnormal{syn}}}{(k_{\textnormal{el}} V_1)}\right)$
that we want to choose.
$x(t,t_0,x_0)$ is the solution to \eqref{eq:sys} with initial
state $x_0$ and initial time $t_0$. There are certain assumptions on the concentrations in the state $x(t)$. It is assumed that $0 \leq x(t) \leq v \text{ for all } t,$
where $v = [v_1, v_2, v_3]^T \in \mathbb{R}^3$ of positive elements (the $v_i$); in this context \enquote{$\leq$} means that the inequality should hold element-wise, and \enquote{$0$} is a vector in $\mathbb{R}^3$ where the elements thereof are equal to zero. It is further assumed that $\frac{R_{\textnormal{syn}}}{(k_{\textnormal{el}} V_1)} \leq u_1(t), \text{ for all } t.$
Without this assumption, the rate $k_c(t)$ could take negative values, which is infeasible. 
The assumptions on $y$ (or rather $f$) lead to different control designs. The matrices $A \in \mathbb{R}^{3 \times 3}$ and $B \in \mathbb{R}^{3 \times 2}$ are defined by 
\begin{align*}
& A =  
 \begin{bmatrix}
-a_{11} & a_{12} & 0\\
a_{21} & -a_{22} & 0 \\
a_{31} & 0 & -a_{33}
\end{bmatrix} = \begin{bmatrix}
\frac{-(k_{12} + k_{\text{el}})}{V_1} & \frac{k_{21}}{V_1} & 0\\
\frac{k_{12}}{V_2} & \frac{-k_{21}}{V_2} & 0 \\
D_{\textnormal{weight}} & 0 & -K_{\textnormal{e}0}
\end{bmatrix}
,& B =  \begin{bmatrix}
1 & 0 & 0 \\
0 & 1 & 0
\end{bmatrix}^T,
\end{align*}
where the $a_{ij}$ are positive for all $i,j$, $D_{weight}$ is the drug molecular weight, $k_{12}$ and $k_{21}$ are the inter-compartmental rates, $V_{2}$ the peripheral volume, and $K_{e0}$ the blood to brain equilibration constant. The matrix $A$
is assumed to be a stable matrix, i.e., all the 
eigenvalues have strict negative real part \cite{lindquist1996introduction,khalil2002nonlinear}.

\section{Comparison of subcutaneous L-dopa infusion}
Using the model in Section~\ref{sec:model}, 
we now compare two different strategies for administration
of levodopa: \textit{constant subcutaneous infusion} and \textit{time-varying subcutaneous infusion based on feedback}. The infusion is only subcutaneous, i.e.,
we design $u_2(t)$ and let the exogenous production term of $u_1(t)$ null for all $t$ (or equivalently $k_c(t) = 0$ for all $t$).

\subsection{Constant infusion}\label{sec:olle1}
The objective considered in here is to find a constant 
rate of infusion $u_2(t)$ such that the concentration
$X_3(t)$ converges asymptotically to a reference value, which is kept within a therapeutic window. Formally, the problem is: 
\begin{problem}\label{prob:1}
Given $r_3 \in \mathbb{R}^+$,
construct a control law $u_2(t)$ such that 
for any any $t_0 \in \mathbb{R}^+$ and $x(t_0) = x_0 \in \mathbb{R}^{3}$,
$X_3(t)$ converges to $r_3$ as $t \rightarrow \infty$.
\end{problem} 
\textbf{Proposed solution to Problem~\ref{prob:1}:} 
\begin{align}
\label{controller:1}
u_2(t) & = -\tilde{r}_2, \quad u_1(t) = 0,
\end{align}
where 
$\tilde{r}_2 = a_{12}r_1 - a_{12}r_2, \:
r_1 = \frac{a_{33}r_3}{a_{31}}, \text{ and } r_2 = \frac{a_{11}r_1}{a_{12}}.$ \\
\noindent 
\textbf{Explanation:}
The proposed solution is very simple and uses
the structure of $A$. It is also robust to small changes
of the parameters of $A$. Let $r = [r_1, r_2, r_3]^T \in (\mathbb{R}^+)^3$
and $[\tilde{x}_1, \tilde{x}_2, \tilde{x}_3]^T = \tilde{x} = x - r$.
It holds that $Ar = [0, \tilde{r}_2, 0]^T$. Now, if we choose $u_2(t)$  as in \eqref{controller:1}, the dynamics for $\tilde{x}$ is $\dot{\tilde{x}} = A\tilde{x},$ which comprises a stable system. Hence,
$\tilde{x}(t)$ converges asymptotically to $r$ when $t$ goes to the infinity,
which means that $X_3(t)$ converges to $r_3$.
Since
there is no feedback, i.e., the control law is constant, there 
is no need to address the issues of time-delays and different time scales in the output. If the input $u$ is slightly off, $x_3(t)$
converges to a value that is close to $r_1$. The most important 
thing to note is that continuous supply of levodopa has the benefit 
of keeping the concentration at a constant reference value, avoiding 
on/off behaviour. 


\subsection{Time-varying infusion and feedback}
In this section, we assume that the reference signal is time-varying. 
Levodopa requirements vary according to foreseeable events, such as circadian rhythm. Patients need less levodopa during the night, although it is important to maintain steady state concentrations to guarantee a better sleeping quality. In this case, the therapeutic threshold is maintained during the day and decreased during night.  High protein diet is also a cause of time-variations in the reference signal: the plasma concentration of levodopa stays stable, while brain concentrations decrease because of competition for blood brain barrier carriers between levodopa and dietary amino acids \cite{nutt1984off}. The structure of the measurements/outputs $y(t)$ determines 
the type of control designs considered.
In this work we will focus on the case when the measurements
of $X_3(t)$ are available -- the concentration 
might be indirectly measured via the effect of hand tapping $E_3(t)$. 

\begin{problem}\label{prob:2}
Given $r_3(t) \in \mathbb{R}^+$,
construct a control law $u_2(t)$ such that 
for any any $t_0 \in \mathbb{R}^+$ and $x(t_0) = x_0 \in \mathbb{R}^{3}$,
$X_3(t)$ converges to $r_3(t)$ as $t \rightarrow \infty$.
\end{problem}

\noindent 
\textbf{Proposed Solution to Problem \ref{prob:2}:} 
\begin{align}
\label{controller:2}
u_2(t) & = -{z(t)}, \quad u_1(t) = 0. 
\end{align}
where $\dot{{z}}(t) = k\tilde{x}_3(t) = k(X_3(t) - r_3)$, $z(t_0) = 0$
and $k > 0$.
The larger the $k$, the 
faster $\|\tilde{x}_3(t)\|$ converges to zero.

\noindent 
\textbf{Explanation:} Now we show that $\|\tilde{x}_3(t)\|$ converges to zero. In the following, all symbols not explained can be found in Section~\ref{sec:olle1}.
We see that $\dot{\tilde{x}} = A\tilde{x} + \tilde{r}_2 - {z(t)}.$
Let $\tilde{z} = \tilde{r}_2 - z(t)$ and 
the augmented state vector be $\tilde{x}_{\text{aug}}(t) = [\tilde{x}(t)^T, \tilde{z}]$. Now one can show that this closed loops dynamics has the 
properties that the $\tilde{x}(t)^T$ vector converges to zero, which means that
$\|\tilde{x}_3(t)\|$ converges to zero. 
Problem~\ref{prob:2} is 
a trajectory tracking problem in a linear system. Since
the system is in strict feedback form one can use  
the backstepping control design technique~\cite{khalil2002nonlinear}. There are also various other methods that can be used.
The proposed controller can be modified in order to account 
for time delays and measurement errors in $X_3(t)$.

\section*{Appendix: Equations of L-DOPA kinetics}
In order to provide the foundations for an L-DOPA control pump (Figure \ref{fig:1}), a comprehensive model is presented combining the drug kinetics with its effects (pharmacodynamics) (Figure \ref{fig:2}). The model consists of two compartments. The first compartment describes the kinetics of L-DOPA following an intravenous (IV) infusion. It also takes into account the synthesis of endogenous L-DOPA. The parameters were identified by fitting the model onto patients' plasma concentrations \cite{chan2005importance}. The second compartment of the model is a result of a study, which focused on the pharmacodynamics of L-DOPA through measuring the hand tapping score (number of taps per minute) with regard to L-DOPA plasma concentration \cite{contin2010pharmacokinetics}. The dynamics were modeled using an $E_{\max}$ model linked to an effect compartment. Patient stratification on Hoehn and Yahr (HY) disease stage allowed to set thresholds for therapeutic efficiency for each stratum. 

    \begin{center}
        \includegraphics[height=6cm]{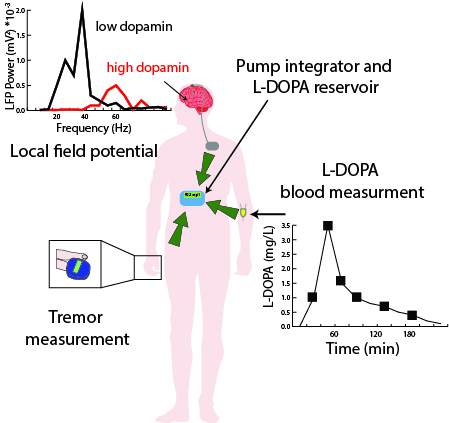}
	\captionof{figure}{Schematic of the closed loop controlled L-DOPA subcutaneous pump.}
	\label{fig:2}  
    \end{center}%
 
\subsection*{Part 1 -- kinetics}
A two-compartmental, patient-validated model is used to model the kinetics of L-DOPA \cite{chan2005importance}.\\
\noindent
\textbf{Compartment 1} (Central compartment).
The continuous infusion case is described by the dynamical model: 
\begin{align*}
\nonumber
\dfrac{dX_1(t)}{dt} & = \dfrac{( - (k_{12} + k_{\text{el}})X_1(t)) + k_{21}X_2(t) + k_c(t)}{V_1} + \frac{R_{\textnormal{syn}}}{(k_{\textnormal{el}} V_1)},
\end{align*}
where $X_1(t)$ is the concentration of L-DOPA in the plasma, $X_2(t)$ the concentration of L-DOPA in the peripheral compartment (see below), and $k_c(t)$ is the rate of infusion in the central compartment: $k_{21}$ and $k_{12}$ are the inter-compartmental rates, $k_{\textnormal{el}}$ is the elimination rate, $R_{\textnormal{syn}}$ is the rate of production of endogenous L-DOPA by the human body, $V_1$ is the volume of central compartment; 
\\
\noindent
\textbf{Compartment 2} (Peripheral compartment).
\begin{equation*}
\dfrac{dX_2(t)}{dt} = \dfrac{(k_{12}X_1(t) - k_{21}X_2(t)) + k_e(t)}{V_2},
\end{equation*}
where $V_2$ is the volume of the peripheral compartment, and $k_e(t)$
is the rate of infusion in the peripheral compartment.

\begin{center}
        \includegraphics[height=6.5cm]{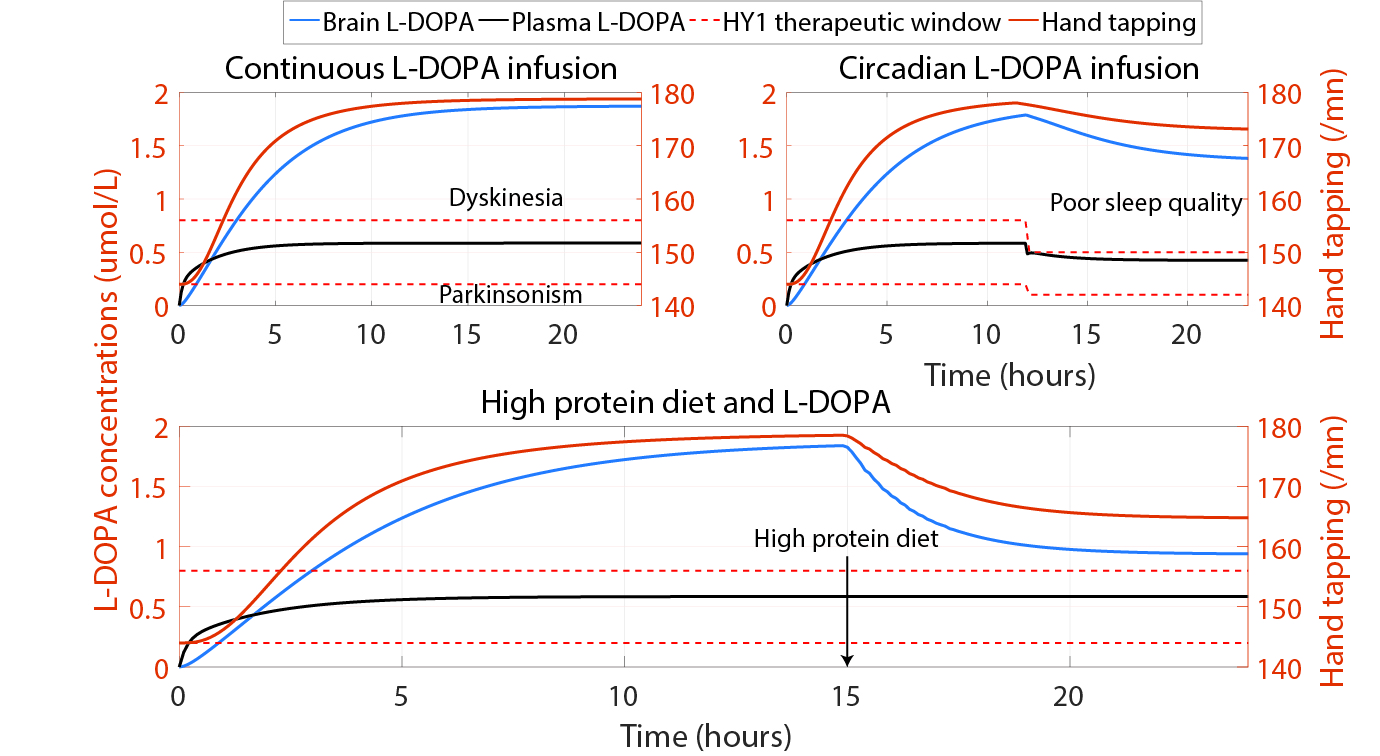}
        	\captionof{figure}{Simulation of L-DOPA kinetics in different settings. (A) Kinetics of L-DOPA in continuous subcutaneous infusion.(B) Cyclic infusion of L-DOPA based on circadian rhythm. (C) Kinetics of brain and plasma L-DOPA in uncontrolled subcutaneous administration with high protein diet, although the plasma concentrations are stable, concentrations in the brain decrease, which provided the rationale for a closed loop subcutaneous infusion to retrieve a profile similar to (A). HY1 refers to the first stage of the disease on Hoehn and Yahr's scale.}
	\label{fig:1} 
\end{center}%

\subsection*{Part 2 -- pharmacodynamics}
The two compartments in Part 1 are complemented 
with an additional compartment -- the \textit{effect compartment} \cite{contin2010pharmacokinetics}. This compartment can be understood as being analogous 
to the brain, where the drug has its effect.\\
\noindent
\textbf{Compartment 3} (Effect compartment).
\begin{equation*}
\dfrac{dX_3(t)}{dt} = X_1(t) - K_{{\textnormal{e}}0}X_3(t),
\end{equation*}
where $X_3(t)$ is the concentration of L-DOPA in the effect compartment, and $K_{{\textnormal{e}}0}$ the equilibration constant that accounts for the delay in reaching the site of action.

\subsubsection*{Acknowledgments}
The authors would like to thank Prof. Thiele for the constructive comments and gratefully acknowledge the financial support by Fonds National de la Recherche (FNR) Luxemourg FNR8864515.


\begin{thebibliography}{4}
\bibitem{contin2010pharmacokinetics} Manuela Contin and Paolo Martinelli. Pharmacokinetics of levodopa.Journal of neurology, 257(2):253–261, 2010.
\bibitem{fahn2015levodopa} Stanley Fahn and Werner Poewe. Levodopa: 50 years of a revolutionary
drug for parkinson disease. Movement Disorders, 30(1):1–3, 2015.
\bibitem{nutt1984off} John G Nutt, William R Woodward, John P Hammerstad, Julie H Carter,
and John L Anderson. The on–off phenomenon in parkinson’s disease:
relation to levodopa absorption and transport. New England Journal of
Medicine, 310(8):483–488, 1984.
\bibitem{schrag2000dyskinesias} Anette Schrag and Niall Quinn. Dyskinesias and motor fluctuations in parkinson’s disease. Brain, 123(11):2297–2305, 2000.
\bibitem{senek2014continuous} Marina Senek and Dag Nyholm. Continuous drug delivery in parkinsons
disease. CNS drugs, 28(1):19–27, 2014.
\bibitem{lundqvist2007continuous} Christofer Lundqvist. Continuous levodopa for advanced parkinsons
disease. Neuropsychiatric disease and treatment, 3(3):335, 2007.
\bibitem{rogers2003elevated} John D Rogers, Anna Sanchez-Saffon, Alan B Frol, and Ramon Diaz-
Arrastia. Elevated plasma homocysteine levels in patients treated with
levodopa: association with vascular disease. Archives of neurology,
60(1):59–64, 2003.
\bibitem{kovatchev2009silico} Boris P Kovatchev, Marc Breton, Chiara Dalla Man, and Claudio
Cobelli. In silico preclinical trials: a proof of concept in closed-loop
control of type 1 diabetes. Journal of diabetes science and technology,
3(1):44–55, 2009.
\bibitem{giladi2015pharmacokinetics} Nir Giladi, Yoseph Caraco, Tanya Gurevitch, Ruth Djaldetti, Yael Cohen,
Oron Yacobi-Zeevi, and Sheila Oren. Pharmacokinetics and safety of
nd0612l (levodopa/carbidopa for subcutaneous infusion): Results from
a phase ii study in moderate to severe parkinsons disease. Age (years),
63(7.4):64–5, 2015.
\bibitem{frankel1989effects} JP Frankel, PA Kempster, M Bovingdon, R Webster, AJ Lees, and
GM Stern. The effects of oral protein on the absorption of intraduodenal
levodopa and motor performance. Journal of Neurology, Neurosurgery
\& Psychiatry, 52(9):1063–1067, 1989.
\bibitem{kleedorfer1991subcutaneous} B Kleedorfer, AJ Lees, and GM Stern. Subcutaneous and sublingual
levodopa methyl ester in parkinson’s disease. Journal of neurology,
neurosurgery, and psychiatry, 54(4):373, 1991.
\bibitem{alonso2006slow} Fortes Alonso-Frech, Ivana Zamarbide, Manuel Alegre, Mar´ıa Cruz
Rodriguez-Oroz, Jorge Guridi, Manuel Manrique, Maria Valencia, Julio
Artieda, and Jos´e Ang´el Obeso. Slow oscillatory activity and levodopainduced
dyskinesias in parkinson’s disease. Brain, 129(7):1748–1757,
2006.
\bibitem{chan2005importance} Phylinda LS Chan, John G Nutt, and Nicholas HG Holford. Importance
of within subject variation in levodopa pharmacokinetics: A 4year
cohort study in parkinsons disease. Journal of pharmacokinetics and
pharmacodynamics, 32(3-4):307–331, 2005.
\bibitem{o2012artificial} Benjamin M O’Brien and Iain Alexander Anderson. An artificial muscle
ring oscillator. Mechatronics, IEEE/ASME Transactions on, 17(1):197–
200, 2012.
\bibitem{lindquist1996introduction} A. Lindquist and J. Sand. An introduction to mathematical systems theory. KTH Lecture Notes, Division of Optimization and Systems
Theory, Royal Institute of Technology (KTH), 1996.
\bibitem{khalil2002nonlinear} H. K. Khalil. Nonlinear systems, volume Third Edition. Prentice hall,2002.
\end{thebibliography}
\end{document}